\documentclass{article}
\usepackage{appendix}
\usepackage{amsmath}
\usepackage[retainorgcmds]{IEEEtrantools}
\usepackage{graphicx}

\begin{document}
\title{Boltzmann \emph{vs} Gibbs: a finite-size match (round 2)}
\author{Loris Ferrari \\ Department of Physics and Astronomy (DIFA) of the University \\via Irnerio, 46 - 40126, Bologna,Italy}
\maketitle
\begin{abstract}
The long standing contrast between Boltzmann's and Gibbs' approach to statistical thermodynamics has been recently rekindled by Dunkel and Hilbert \cite{DH}, who criticize the notion of negative absolute temperature (NAT), as a misleading consequence of Boltzmann's definition of entropy. A different definition, due to Gibbs, has been proposed, which forbids NAT and makes the energy equipartition rigorous in arbitrary sized systems. The two approaches, however, are shown to converge to the same results in the thermodynamical limit. A vigorous debate followed ref.~\cite{DH}, with arguments against \cite{FW, Ual} and in favor \cite{DH', DH'', C, S} of Gibbs' entropy. In an attempt to leave the speculative level and give the discussion some deal of concreteness, we analyze the practical consequences of Gibbs' definition in two finite-size systems: a non interacting gas of $N$ atoms with two-level internal spectrum, and an Ising model of $N$ interacting spins. It is shown that for certain measurable quantities, the difference resulting from Boltzmann's and Gibbs' approach vanishes as $N^{-1/2}$, much less rapidly than the $1/N$ slope expected. As shown by numerical estimates, this makes the \emph{experimental} solution of the controversy a feasible task.\newline       
\end{abstract}

e-mail: loris.ferrari@unibo.it
telephone: ++39-051-2091136\newline
\\
\textbf{Preamble}

A first version of the present work was delivered to ArXiv months ago, and, jointly, to \emph{The Physical Review E}. After a long process, the manuscript was finally rejected, due to the negative opinion of one referee (the other was, istead, very favorable), concerning the application of Gibb's entropy to Weiss ferromagnetism.

ArXiv is an open space, where people can freely show their results and opinions, leaving any judgement to a free audience of readers. Hence, it seems to the author that the above piece of information is important: the reader is now acquainted that there was a contrast, on the content of the work, and, after a reading, she/he can judge about the terms of the contrast, and decide which opinion is right (if any: comments are welcome). Actually, sometimes the debate raised by a work is more fruitful than the work itself.

At the end of the work, a post-amble is added, in which the reasons of the referee's negative judgement are reported (as far as the author did understand them).

\section{Introduction}
The notion of negative absolute temperature (NAT) \cite{PP,R} has been recently refreshed by Braun et \emph{al} \cite{Bal} . The main factor of novelty was the claim that the NAT regime was attained in a system with a \emph{continuous} upperly bounded spectrum, in contrast to the preceding experiments based on two-level systems \cite{PP,R}. Soon after, however, the notion of NAT was criticized by Dunkel and Hilbert \cite{DH}, as a misleading consequence of Boltzmann's definition of entropy
\begin{equation}
\label{Sb}
S_B\left(E\right)=\mathrm{ln}\left[\sum_{\eta}\delta_{H\left(\eta\right),\:E}\right]\:,
\end{equation}
in an isolated (microcanonical) system of Hamiltonian $H(\eta)$, $\eta$ being any state variable and $\delta$ a generalized Kronecher symbol (Boltzmann constant $k_B=1$). Actually, from the definition of absolute temperature
\begin{equation}
\label{Tdef}
T=\left(\frac{\partial S}{\partial E}\right)_V^{-1}\:,
\end{equation}
it is seen that NAT occurs whenever $S(E)$ is a \emph{decreasing} function of the energy, which is possible for the entropy \eqref{Sb}, if the energy is upperly bounded (i.e. $H(\eta)<E_{Max}$) and the higher energy levels can be overpopulated by suitable external processes. As stressed in Ref. \cite{DH}, instead, Gibbs' entropy
\begin{equation}
\label{Sg}
S_G\left(E\right)=\mathrm{ln}\left[\sum_\eta \Theta\left(E-H\left(\eta\right)\right)\right]
\end{equation}
($\Theta(\cdot)$ being the Heaviside function) is an increasing function of $E$ and thereby excludes any NAT regime by definition. In Ref. \cite{DH} the authors stress that Gibbs' definition of entropy refers to a microcanonical system (a claim that will be reconsidered in what follows) and satisfies the equipartition theorem in any case, while Boltzmann's entropy fails in very small systems, with a number of particles of order unity.

What we playfully call the "Boltzmann \emph{vs} Gibbs match"\footnote{This title should not be confused with Jaynes' one~\cite{J}: \emph{Gibbs} vs \emph{Boltzmann entropies}, that deals with the (supposed) inadeguacy of Boltzmann combinatorial method in describing \emph{interacting} systems.} is right the debate about Boltzmann's (eq.n~\eqref{Sb}) and Gibbs' (eqn.~\eqref{Sg}) definition of entropy \cite{FW, Ual, DH', DH'', C, S}. As it is a rule for most polemics involving the entropy, the discussion may look a little bit academic, especially in view of equation~(14) of ref. \cite{DH}, which shows that the two entropies \eqref{Sb} and \eqref{Sg} give the same temperature expression, in the thermodynamic limit, with a difference that vanishes as the inverse heat capacity. Given that Boltzmann and Gibbs's picture lead to the same results in large systems, the alternative looks as follows: either there exists a \emph{measurable size effect} displaying some difference between eqn.s~\eqref{Sb} and \eqref{Sg}, or any discussion remains confined to a merely speculative level. The second possibility looks, at a first sight, much more likely, once one efforts the question how small the size should be, for this effect to be detectable. Actually, one might be tempted to assume that the smallness criterion is determined by $1/N$ ($N$ being the number of particles), since from eq.n~(14) of ref.~\cite{DH} the temperatures resulting from eq.n~\eqref{Sg} and \eqref{Sb} differ by terms proportional to the inverse of an extensive quantity (the heat capacity). If this were always the case, any experimental test, like one based on the "minimal quantum thermometer" suggested in ref.~\cite{DH}, would become extremely difficult, if not impossible. Fairly surprisingly, we shall show that, in certain two-level systems, the difference between certain measurable quantities, derived from eq.n~\eqref{Sg} and \eqref{Sb}, vanishes as $1/\sqrt{N}$. This result makes the experimental comparison between the two approaches much more accessible, since detecting effects small to order $1/\sqrt{N}$ is certainly easier than to order $1/N$, especially in view of the recent technical progresses of small particle physics. This is the "finite-size match" that we are going to outline in the next sections.

Though we shall refer to two-level systems, that represent the preferred framework for NAT experiments, we are not specifically interested in the NAT problem. We address to a more fundamental point, that intrigues people since the foundations of statistical thermodynamics, i.e. the definition of entropy in terms of the microscopic dynamical states of the system. Very recently, Hilbert, H\"anggi and Dunkel performed an exhaustive survey of several possible definitions of entropy in continuous spectrum systems \cite{HHD}, with special attention to their consistency with the three principles of thermodynamics. It turns out that the only definition of entropy that satisfies the three principles in arbitrary small systems, is eq.n~\eqref{Sg}. The flaws resulting from other definitions are, in turn, finite-size effects whose \emph{measured} presence would support a widespread opinion (contrasted by Hilbert, H\"anggi and Dunkel) that thermodynamics applies only to large systems.  

Another point that we address in the present work is the role of the constraints in the construction of the thermodynamical functions. Actually, different constraints (microcanonical, canonical, grandcanonical) yield different fluctuations about the equilibrium values of the thermodynamical quantities. Another widespread opinion is that the equilibrium values are, themselves, independent from the constraints. This justifies the experimental evidence that the constraints are irrelevant in the thermodynamical limit (and far from the phase transitions), since the relative weights of the fluctuations vanish as $1/\sqrt{N}$. However, the difference between eq.n~\eqref{Sg} and \eqref{Sb} is not due to the fluctuations, but to the genuine notion of thermal equilibrium. Indeed, Gibbs's entropy \eqref{Sg} postulates what, in Boltzmann's language\footnote{Which is hated and rejected by many Gibbs supporters~\cite{Gull}.}, sounds like a sort of new ergodic hypothesis, that (in contrast to what claimed in ref.~\cite{DH}) looks hardly appliable to micro-canonical systems. Why should states with energy \emph{less} than $E$ be involved, if $E$ is strictly conserved? In this concern, it is only Boltzmann's entropy \eqref{Sb} that is consistent with the dynamics of an isolated system. A reasonable possibility is that Gibbs' and Boltzmann's entropies refer to \emph{canonical} and \emph{micro-canonical} constraints, respectively, which is right the opposite of what the supporters of Gibbs' entropy do claim \cite{DH, DH', C}. The difference between canonical and microcanonical systems will come into play in the study of Weiss ferromagnets (Section~\ref{Weiss}). All the way, it leaves a relevant question pending, to which finite-size experiments could give an answer.

What precedes motivates our program of giving the size effects of interest an analytical form, then studying the feasibility of \emph{ad hoc} experiments revealing which of the two entropy definitions does fit better with physical reality, both in canonical and in micro-canonical systems. In order to avoid any possible factor of confusion, it is worthwhile stressing the strategy used in the next sections. We calculate the quantities of interest as functions of the number $N$ of particles, according to Gibbs' definition eq.n~\eqref{Sg}. Then we compare such quantities with those obtained according to Boltzmann's definition \eqref{Sb}, under the \emph{same} condition of large but finite $N$. In Section~\ref{Weiss} (the Ising ferromagnet) this procedure is manifest (see, for example, Fig.s 2, 3 and 4). In Section \ref{2levgas}, the Boltzmann-formulated quantity (the energy of the two-level gas) is reported from other papers (in particular, Ref.~\cite{DH}). 

Another \emph{caveat} is in order: the current way to represent the differences between Boltzmann's and Gibbs' entropies is writing the Boltzmann/ Gibbs "temperatures" $T_{B,G}(\eta)$, as functions of other state parameter(s) $\eta$. However, this does not mean introducing different  Boltzmann/Gibbs "thermometers", but different relationships between the \emph{measured} temperature $T$ and $\eta$. Right to point out that $T$ is the \emph{same measured temperature}, both for Boltzmann and for Gibbs, we  will not speak about Boltzmann's or Gibbs' "temperature" (apart from rare cases) and will not append any subscript G or B to $T(\eta)$.

\section{A gas of non interacting two-level particles}
\label{2levgas}
The simplest system for testing the differences between the two entropies \eqref{Sb} and \eqref{Sg} is a non interacting gas of $N$ two-level particles, with populations $n_\pm$ in each level $\pm\epsilon$. While Boltzmann predicts $n_+=n_-=N/2$, for $T\rightarrow\infty$\footnote{We implicitly assume that the two levels are non degenerate or have the same degeneration.}, Gibbs predicts $n_+(T)$ increasing continuously up to an inverted population regime $n_+\rightarrow N,\:\:n_-\rightarrow0$ at $T=\infty$. At this stage one might wonder why laser devices need a resonant external field, in order to get an overpopulated level, if this could be obtained simply by thermal activation. The point is that in the Gibbs framework, the \emph{size} of the gas (represented by the particle number $N$) plays a crucial role in determining how high the temperature should be, to achieve a significant overpopulation of the upper level. This is what we are going to show in the present section.

On setting $n_{\pm}=(N\pm m)/2$, the energy of the gas (as far as the two-level part is concerned) reads:
\begin{equation}
\label{E1}
\mathcal{E}(m)=\epsilon\;m\:.
\end{equation}
Since the condition $\mathcal{E}(m)<E$ implies $m<E/\epsilon$, Gibbs' entropy reads:
\begin{equation*}
S_G\left(E\right)=\mathrm{ln}\left[\sum_{m\le E/\epsilon}\frac{N!}{\left(\frac{N+m}{2}\right)!\left(\frac{N-m}{2}\right)!}\right]\:.
\end{equation*}
If $N\gg1$, the entropy can be expressed in an integral form, making use of Stirling formula for the factorials:
\begin{subequations}
\label{Sg1,phi,z}
\begin{equation}
\label{Sg1}
S_G\left(E\right)=\mathrm{ln}\left[2^N\int_{-1}^{z(E)}\mathrm{d}x\:e^{-\frac{N}{2}\phi(x)}\right]\:,
\end{equation}
with
\begin{equation}
\label{z}
z(E)=\frac{E}{N\epsilon}\in[-1,1]
\end{equation}
and
\begin{align}
\phi(x)&=\left[(1+x)\mathrm{ln}(1+x)+(1-x)\mathrm{ln}(1-x)\right]=\nonumber\\
&=\phi(z)+\underbrace{\mathrm{ln}\left(\frac{1+z}{1-z}\right)}_{a(z)}(x-z)+\underbrace{\frac{1}{1-z^2}}_{b(z)}(x-z)^2+\dots
\label{phi}
\end{align}
\end{subequations}
The series expansion in eq.n~\eqref{phi} will be used in Appendix A. The temperature follows from eq.n~\eqref{Tdef}, on account of eq.ns~\eqref{Sg1,phi,z}: 
\begin{equation}
\label{Tgas1}
T(E)=N\epsilon\;e^{\frac{N}{2}\phi\left(z(E)\right)}\int_{-1}^{z(E)}\mathrm{d}x\:e^{-\frac{N}{2}\phi(x)}\:.
\end{equation}
For $N\gg1$, the following expressions are derived in Appendix A:
\begin{subequations}
\label{E< >0}
\begin{align}
\frac{T(E)}{2\epsilon}&=\left[\mathrm{ln}\left(\frac{1-z(E)}{1+z(E)}\right)\right]^{-1}& \text{for } E<0,\; |E|\gg\epsilon\sqrt{\frac{N}{2}}\label{E<0}\\
& \nonumber \\
&=\frac{1}{2}\sqrt{\frac{\pi N}{2}} & \text{for } E=0 \label{E=0}\\
& \nonumber \\
&=\frac{1}{2}\sqrt{\frac{\pi N}{2}}\mathrm{exp}\left[\frac{N}{2}\phi\left(z(E)\right)\right]\times&\nonumber\\
&\times\left[1+\mathrm{erf}\left(z(E)\sqrt{\frac{N}{2}}\right)\right] & \text{for } 0<E\ll\epsilon\:N\label{E>0}\:.
\end{align}
\end{subequations}
The condition $E<0$ corresponds to the thermal regime in which the Boltzmann temperature $T_B$ is positive and the lower level is always more populated then the higher one. Actually, equation \eqref{E<0} recovers exactly the standard result: 
\begin{equation}
\label{zb}
z(T)=\frac{e^{- \epsilon/T}-e^{\epsilon/T}}{e^{\epsilon/T}+e^{- \epsilon/T}}\:,
\end{equation}
as can be easily shown by solving \eqref{E<0} with respect to $z(E)$, then using eq.ns~\eqref{z} and \eqref{E1}. The differences between Boltzmann and Gibbs at finite $N$ becomes divergingly large in the range of critical energies: 

\begin{equation}
\label{DeltaE}
-\underbrace{\epsilon \sqrt{N/2}}_{\Delta E}< E<0\:.
\end{equation}

Indeed, it  is well known that $\mathrm{lim}_{E\rightarrow0^{-}}T(E)=\infty$, according to Boltzmann, for any value of $N$ (see, for instance, Fig. 1 in ref.~\cite{DH}). In contrast, equation~\eqref{E=0} shows that $\lim_{E\rightarrow 0^-}T(E)=T_G\propto\sqrt{N}$ is \emph{finite}, but diverging with $\sqrt{N}$, according to Gibbs. The $1/\sqrt{N}$-discrepancy between Boltzmann and Gibbs is expressed by the ratio $\Delta E/(\epsilon N)$, between the range of critical energies (eq.n~\eqref{DeltaE}) and the total energy range. Note that the condition $E>0$ corresponds to what would be the NAT regime, that should be attained by "super heating" the system above $E=0$, according to Boltzmann's picture\footnote{This is what Sokolov criticizes as "hotter than hot" \cite{S}.}. 

From a formal viewpoint, the results eq.ns~\eqref{E< >0}, obtained in the Gibbs framework, cure the singularity of the Boltzmann temperature at $E=0$ and bring the gas continuously to any positive value of the energy, i.e., to any \emph{thermally activated} overpopulation of the upper level. However, it is easily shown from eq.n~\eqref{E>0} that the temperature diverges \emph{exponentially} with $N$, for $E>0$ (Fig. 1), while the crossing temperature
\begin{figure}[htbp]
\begin{center}
\includegraphics[width=4in]{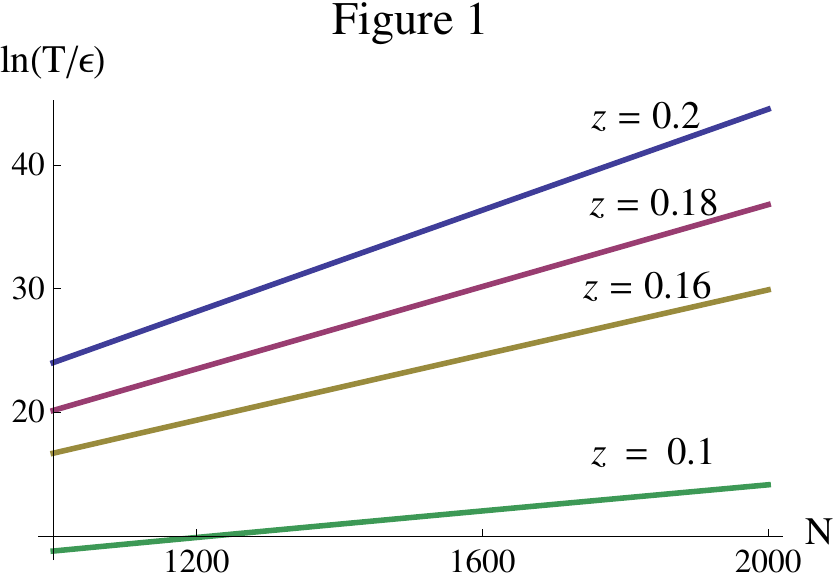}
\caption{\textbf{Gibbs temperature of a two-level gas in the Boltzmann NAT regime $\mathbf{E>0}$.}\newline
The logarithmic plot shows the exponential increase of $T$ in the particle number $N$, for different \emph{positive} values of $z$, corresponding to a \emph{thermally activated} overpopulation of the upper level.}
\label{default}
\end{center}
\end{figure}

\begin{equation}
\label{TG}
T_G=\epsilon\sqrt{\frac{\pi N}{2}}
\end{equation} 
at $E=0$ (eq.n~\eqref{E=0}), diverges with the square root of the system size. Hence, the regime $E>0$ would be unaccessible to any \emph{thermal} process, for $N\rightarrow\infty$, even in the Gibbs framework. In this (non trivial) sense Gibbs and Boltzmann do converge to the same results in the thermodynamic limit. However, in a finite-size system, the regime $E>0$ (or, equivalently, $z>0$) is not forbidden in principle, according to Gibbs, and could be accessed by an appropriate experimental set up. This is what we are going to exploit in what follows.

Let us derive an expression for $z$ as a function of $T$ and $N$, recalling that $z$ (eq.n~\eqref{z}) is the fractional difference between the higher and lower level populations. Assuming $0<z\ll1$, which means $\phi(z)=z^2+\cdots$, and defining $\theta=z\sqrt{N/2}$, equation~\eqref{E>0} yields:
\begin{equation}
\label{z2}
\theta=\left[\mathrm{ln}\left(\frac{T}{T_G\left(1+\mathrm{erf}(\theta)\right)}\right)\right]^{1/2}\;\Rightarrow\; z\approx\sqrt{\frac{2}{N}\mathrm{ln}\left(\frac{T}{T_G}\right)}\quad(T>2\:T_G)\:,
\end{equation} 
since $\mathrm{erf}(\theta)$ ranges between $0$ and $1$ and does not affect the result significantly\footnote{The condition $T>2\:T_G$ ensures that the quantity in square root is positive, for any value of $\theta$, but the solution of the equation exists for any $T>T_G$. Note that $z\ll1$ does \emph{not} imply $\theta\ll1$.}. At fixed temperature $T$ ($>2\:T_G$), the overpopulation factor $z$ decreases with the square root of the system size. Due to the logarithmic dependence, the value of $z$ is almost insensitive to $T$. Hence, the feasibility of an experiment possibly measuring $z>0$ at a \emph{positive} temperature rests on two main difficulties: first, a crossing temperature $T_G\propto\sqrt{N}$ not so large as to prevent high precision measurements; second, an overpopulation factor $z\propto\sqrt{1/N}$ not so small as to escape the instrumental sensitivity.

Consider a gas of $N$ atoms of $^{3}\mathrm{He}$, in a volume $V$, each carrying a spin magnetic moment $\mu_{Bohr}\approx 10^{-20}\mathrm{emu}$. Let $B$ be an external uniform magnetic field. From eq.n~\eqref{z2}, the total induced moment $\mu_{tot}=\pm\mu_{Bohr}\left(n_+-n_-\right)$ reads, according to Gibbs:
\begin{subequations}
\begin{equation}
\mu_{tot}^G\approx-\mathrm{sign}\left(B\right)\mu_{Bohr}\sqrt{2N\mathrm{ln}\left(\frac{T}{T_G}\right)}\:,\label{mugasG}
\end{equation}
while Boltzmann (eq.n~\eqref{zb} with $\epsilon\ll T$) yields:
\begin{equation}
\label{mugasB}
\mu_{tot}^B\approx \mathrm{sign}\left(B\right)\mu_{Bohr}N\frac{\epsilon}{T}\:,
 \end{equation}
 \end{subequations}
with $\epsilon=\left|B\right|\mu_{Bohr}$. Note that $\mu_{tot}^G$ is \emph{antiparallel} and $\mu_{tot}^B$ \emph{parallel} to $B$, since the former comes from the overpopulation of the upper level, while the latter comes from the overpopulation of the lower one. SQUID magnetometers, that are the highest sensitivity instruments currently available, can measure magnetic moment intensities down to $\mu_m\approx10^{-8}\mathrm{emu}$. For $T$ larger than, but comparable to $T_G$, the condition $\left|\mu_{tot}^G\right|\approx \mu_m$ (eq.n~\eqref{mugasG}) yields the order of magnitude $N\approx10^{24}$ for the minimum number of atoms required to produce a detectable value of $\mu_{tot}^G$. Efficient SQUID devices usually operate at temperatures of $10\:\mathrm{K}$ at most, which determines an upper limit for $T_G$. From eq.n~\eqref{TG}, however, this turns into an upper limit for 
\begin{equation}
\label{Bmin}
\left|B\right|=\sqrt{\frac{2}{N\pi}}\frac{T_G}{\mu_{Bohr}}\approx10^{-7}\mathrm{G}\:.
\end{equation}
With those values of $N$, $B$ and $T$, one sees that $\left|\mu_{tot}^B\right|\approx10^{-8}\mathrm{emu}$ (eq.n~\eqref{mugasB}) is comparable in magnitude to $\left|\mu_{tot}^G\right|$, so that the measurable difference between Boltzmann's and Gibbs' predictions results mainly in the \emph{opposite} orientation of a small induced magnetic moment. This effect is $1/\sqrt{N}$-small because the \emph{antiparallel} induced magnetic moment $\mu_{tot}^{G}$ increases as $\sqrt{N}$ (eq.n~\eqref{mugasG}), while the \emph{parallel} induced moment $\mu_{tot}^{B}$ is proportional to $N$ (eq.n~\eqref{mugasB}).

As for the volume $V$ of the gas, one should recall that the size of the sample does also play a role, in the feasibility of the experiment. It can be easily seen that $V\approx10\:\mathrm{cm}^3$, corresponding to a density of $10^{23}\mathrm{cm}^{-3}$ of $^3\mathrm{He}$ atoms, at $T\approx10\:\mathrm{K}$, yields degeneration effects less than $10\%$. This looks a still tolerable error factor for the implicit approximation assumed so far, that the two-level energy is statistically independent from the translational energy (which is false, for degenerate gases).        

In conclusion, some cube centimeters of $^3\mathrm{He}$, with density about $10^{23}\mathrm{cm}^{-3}$, at a temperature about $10\:\mathrm{K}$, under the action of a magnetic field of $10^{-7}\mathrm{G}$, could be a good candidate as a referee of the Boltzmann \emph{vs} Gibbs match. If the measured magnetic moment (spin polarization) of the gas, induced by the field, were \emph{anti-parallel} to the field itself, the resulting  \emph{thermally activated} overpopulation of the upper level would provide a strong experimental support to Gibbs' definition of entropy. On suitably changing the experimental set up, the hypothesis that Gibbs' and Boltzmann's entropies do refer to different constraints (microcanonic or canonic) could be exploited in turn.

\section{Weiss ferromagnetism revisited}
\label{Weiss}

Another system for testing the consequences of Gibb's definition of entropy is a ferromagnetic material, modeled by an Ising lattice of $N=n_++n_-$ interacting magnetic moments with only "up" ($+$) and "down" ($-$) orientations. In this case one has to face the problem of applying Gibbs' picture to a phase transition (paramagnetic $\leftrightarrow$ ferromagnetic), which turns out to be a non trivial issue. 

In the mean field approximation, the energy of the Ising model reads:
\begin{equation}
\label{E2}
\mathcal{E}(z,\:\rho)=-JN\left(z^2+2\rho\: z\right)\:,\;z=\frac{n_+-n_-}{N}\:,
\end{equation}  
where $J>0$ is half the coupling constant of the moment-moment interaction, $z$ is the magnetization and $\rho\equiv \mu_0B/(2J)$ is the energy contribute from an external uniform magnetic field, in units of $2J$. In a  first-order approximation in $\rho$ ($|\rho|\ll1$), the condition $\mathcal{E}(x,\:\rho)\le E$ implies:
\begin{equation*}
x\ge |z(E)|-\rho\quad \text{or}\quad x\le -|z(E)|-\rho\:,
\end{equation*}
with
\begin{equation*}
|z(E)|=\left|\frac{E}{JN}\right|^{1/2}\quad \text{and}\quad-\left(1+\rho\right)\le\frac{E}{JN}\le0\:.
\end{equation*}
In the same limit $N\gg1$ and with the same method used for eq.n~\eqref{Sg}, the preceding formulas yield, for Gibbs' entropy:
\begin{align}
S_G(z)&=\mathrm{ln}\left[2^{N}\left(\int_{|z|-\rho}^{1}\mathrm{x}\:e^{-\frac{N}{2}\phi(x)}+\int_{-1}^{-|z|-\rho}\mathrm{d}x\:e^{-\frac{N}{2}\phi(x)}\right)\right]=\nonumber\\
&=\mathrm{ln}\left[2^{N+1}\int_{|z|}^{1}\mathrm{d}x\:e^{-\frac{N}{2}\phi(x)}+\boldsymbol{\circ}\left(\rho^2\right)\right]\:.\label{Sg2}
\end{align}

\begin{figure}[htbp]
\begin{center}
\includegraphics[width=4in]{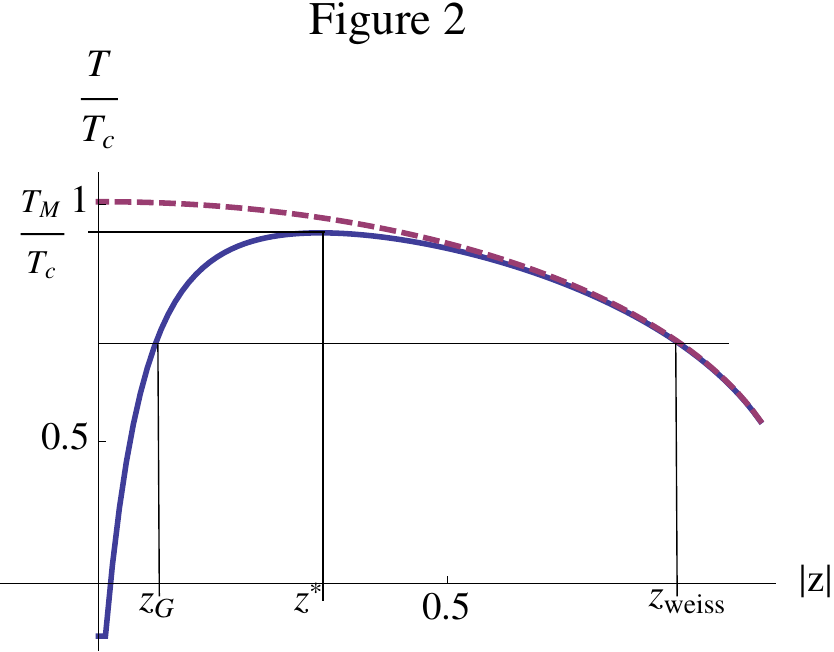}
\caption{\textbf{Gibbs' and Boltzmann's temperatures of a Weiss ferromagnet (Ising model) in zero external field ($\mathbf{N=200}$).}\newline
Dashed and full lines refer to Boltzmann's and Gibbs' temperature, respectively, as functions of the relative magnetization intensity $|z|$. For the meaning of $z_{weiss}$, $z^*$ and $z_G$, see the text.}
\label{default}
\end{center}
\end{figure} 

Let the external magnetic field vanish ($\rho=0$). On applying eq.n~\eqref{Tdef} to eq.ns~\eqref{E2} and \eqref{Sg2}, the relationship between Gibbs' temperature and the relative magnetization intensity $|z|$ is shown by the full line in Fig. 2. The dashed line, instead, refers to Boltzmann's temperature. The right side of Fig. 2 shows that both $T$-curves initially increase with decreasing $|z|$, which is what one expects for a ferromagnet whose spontaneous magnetic order tends to be destroyed by an increasing thermal disorder. Boltzmann's temperature increases on and tends to the Curie temperature $T_c=2J$  for $|z|\rightarrow0$. Gibbs' temperature, instead,  attains a maximum $T_M$, lower than $T_c$, then \emph{decreases with decreasing} $|z|$, down to zero, at $|z|=0$ (left side of Fig. 2). A common drawback of both curves is that they are unable to account for the paramagnetic phase at $T>T_c$. In addition, Gibbs' temperature exhibits further intriguing aspects: for each $T<T_M$, there are \emph{two} possible values $z_{weiss}$ and $z_G$ of the relative magnetization intensity. Since $\partial T/\partial z$ is positive in $z_G$ (and negative in $-z_G$), according to equation~\eqref{E2}, it is easy to show that the specific heat at constant volume is \emph{negative} in $\pm z_G\footnote{In the next formulas we use the following convention: $\left[\partial A/\partial\eta\right]_{\eta_{eq}}$ means: derivative of quantity $A(\eta)$, calculated in $\eta_{eq}$;  $\left(\partial B/\partial X\right)_{V,\cdots}$ means: derivative of quantity $B$ with respect to $X$, keeping $V,\cdots$ constant.}$:

\begin{equation}
\label{CG}
C_{G}=\left[\left(\frac{\partial\mathcal{E}}{\partial z}\right)_V\left(\frac{\partial T}{\partial z}\right)^{-1}\right]_{\pm z_G}<0\:.
\end{equation}

Furthermore, $\partial\mathcal{E}/\partial|z|$ is \emph{finite} at $|z|=z^*$, in which the derivative of $T(|z|)$ \emph{vanishes}. Hence:

\begin{equation}
\label{lim}
\mathrm{lim}_{|z|\rightarrow (z^*)^\pm}\frac{\partial\mathcal{E}}{\partial|z|}\left(\frac{\partial T}{\partial|z|}\right)^{-1}=\pm\infty\:.
\end{equation}

Equation~\eqref{lim} shows that the specific heat $C_G$ diverges positively or negatively for $T\rightarrow T_M$, depending on which state, $z_{weiss}$ or $z_G$, is adopted. Though a phase transition may well produce a multiplicity of equilibrium states at the same $T,\:V,\cdots$, in this case it is clear that the "states" $\pm z_G$ are spurious and that the range of values $0<|z|<z^*$ in Fig. 2 is unphysical in some sense. Here we suggest an argument that provides a plane interpretation of Fig. 2, but challenges the position that Gibbs' entropy and temperature refers to \emph{microcanonical} systems only, while Boltzmann's picture refers to \emph{canonical} systems \footnote{In ref.~\cite{C}, for example, it is shown that Gibbs' entropy is the unique expression that fits rigorously with Clausius definition of temperature in a microcanonical system. When dealing with a canonical system, instead, it is Shannon's entropy (Shannon's and Boltzmann's entropy coincide, in a canonical system) that shares the same property. In ref.\cite{DH'} it is stressed that the difference between $S_B$ and $S_G$ could result right from the difference between canonical and microcanonical constraints (See footnote [6] in ref.~\cite{DH'})}. On assuming that eq.n~\eqref{Tdef} does apply, at least locally, one has:  

\begin{equation}
\nonumber
T=\left(\frac{\partial\mathcal{E}}{\partial S}\right)_V=\left[\frac{\left(\partial\mathcal{E}/\partial\eta\right)_V}{\left(\partial S/\partial\eta\right)_V}\right]_{\eta_{eq}}\:,
\end{equation}

for any state parameters $\eta$ that characterizes the system, and attain the equilibrium values $\eta_{eq}$. The next step follows from rephrasing the preceding formula as:

\begin{equation}
\label{StabCond2}
\left(\left[\frac{\partial\left(\mathcal{E}-TS\right)}{\partial\eta}\right]_{\eta_{eq}}\right)_{V,T}=0\:,
\end{equation}

which yields the extremants of the Helmholtz free energy $\Psi=\mathcal{E}-TS$. In Appendix 2, it is shown that the spurious states $\pm z_G$ correspond to \emph{maxima} of the Helmholtz free energy $\Psi_G=\mathcal{E}-TS_G$, while the state $z=0$ is always a cuspid minimum, i.e. $\Psi_G(z)-\Psi_G(0)= T|z|\sqrt{2/(N\pi)}$ for small $|z|$ (eq.n~\eqref{cuspid}). However, this makes sense only if the temperature $T$, defined through eq.n~\eqref{Tdef}, refers to a \emph{canonical} system, for which the minimization of  the Helmholtz free energy is the correct approach to thermal equilibrium. Hence, the interpretation just outlined stems from the assumption that Gibbs' entropy does apply to \emph{canonical} systems too (or only), which is a pending question. Leaving the alternative between microcanonical or canonical use of $S_G$ to future discussions and investigations, we take the states $z_{weiss}$ (with positive specific heat) as the ones corresponding to the true thermal equilibrium, and stress the most relevant result following from Fig. 2:

\begin{figure}[htbp]
\begin{center}
\includegraphics[width=4in]{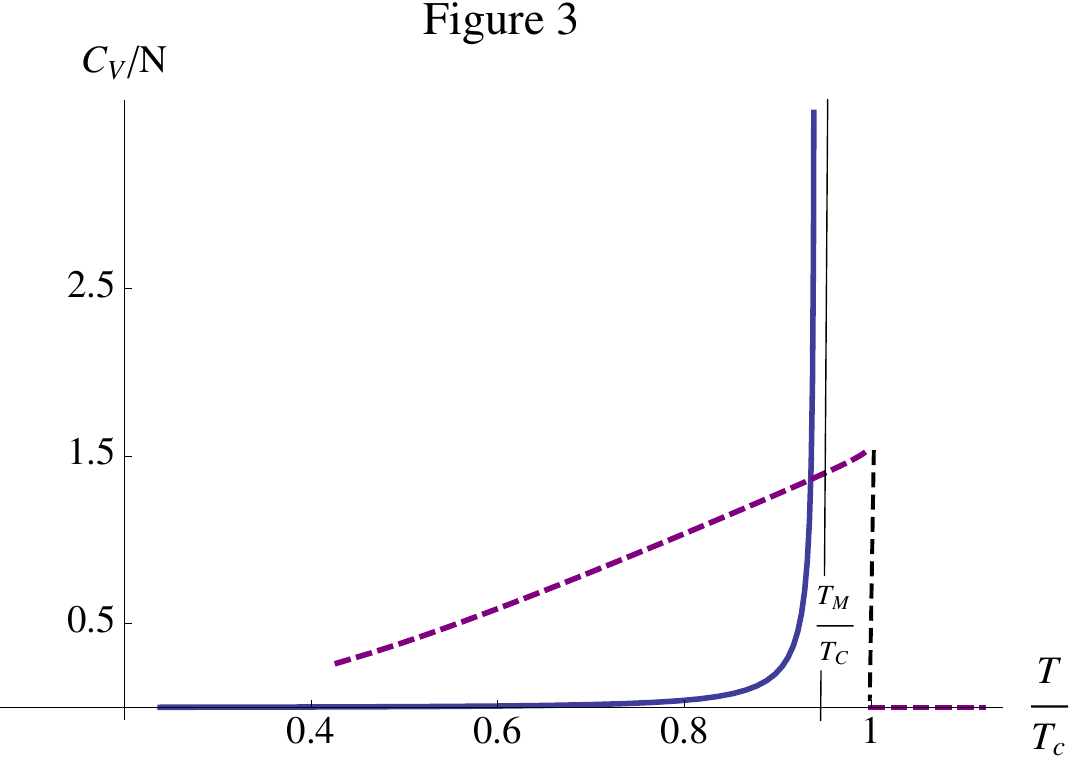}
\caption{\textbf{Gibbs and Boltzmann heat capacities as a function of $T$ in a Weiss ferromagnet ($N=200$).}\newline 
A finite discontinuity at $T_c=2J$ (dashed line) \emph{vs} a divergent discontinuity at $T_M<T_c$ (full line) characterize Boltzmann's and Gibbs' heat capacity, respectively.}
\label{default}
\end{center}
\end{figure}

Figure 3 shows the behavior of Boltzmann's (dashed line) and Gibbs' (full line) heat capacities: while the former displays a \emph{finite} discontinuity at the Curie temperature $T_c=2J$ (which is reported even in the pedagogical literature \cite{KPW}), the latter exhibits a \emph{divergent} discontinuity at the critical temperature $T_M$, lower than $T_c$. 

As far as our aims are concerned, the question turns back to the possible experimental evidence of the diverging heat capacity mentioned above. At a first sight, this effect looks huge and very easy to observe. Indeed, one might wonder why it has not been reported before, if Gibbs is right. The problem is much less trivial, actually. First, it can be seen that $|T_c-T_M|\propto 1/\sqrt{N}$, i.e. the difference between the standard Curie temperature and Gibbs-Curie temperature $T_M$ is a size effect, vanishing as the inverse square root of the system size (see Fig. 4). This means that the temperature range in which the diverging behavior could be observed vanishes in turn, with at least the same $1/\sqrt{N}$-slope. In addition, under the hypothesis that Gibbs' picture is \emph{only} microcanonical, the $C_B$ divergence could be observed only if the phase transition is approached in isolation. In ref.~\cite{C}, the practical difficulty of implementing such a process is stressed with some interesting details. So, the experimental check of Fig. 3 turns out to be far from easy. If, instead, Gibbs' picture were also canonical, Section~\ref{Weiss2} outlines a possible experimental route, based on magnetic measurements, that turns out to be feasible, with appropriate high-level instruments.

\begin{figure}[htbp]
\begin{center}
\includegraphics[width=4in]{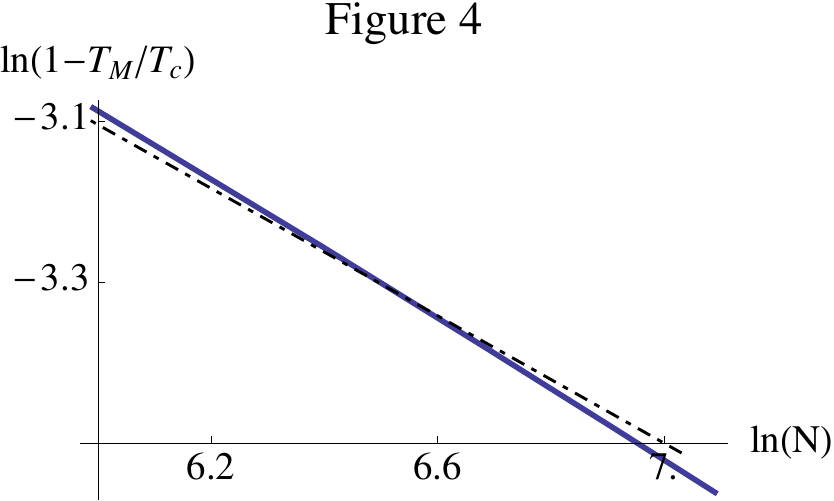}
\caption{\textbf{Log-Log plot of $1-T_M/T_c$ as a function of the system size $N$}.\newline
The numerical data (full line) are compared to the predicted $1/\sqrt{N}$ slope (dot-dashed line).\newline}
\label{default}
\end{center}
\end{figure}

\section{Weiss Ferromagnetic transition as a canonical process}
\label{Weiss2}

As anticipated in Section~\ref{Weiss}, here we assume that both Boltzmann's and Gibbs' entropies apply to \emph{canonical} systems, and thereby proceed to minimize the free energies

\begin{equation}
\label{Psialpha}
\Psi_\alpha(z)=\mathcal{E}(z)-TS_\alpha(z)\quad(\alpha = B,\:G)
\end{equation}

with respect to the magnetization $z$, according to eq.n~ \eqref{StabCond2}. For the sake of brevity, we do not report the explicit formulas for $\Psi_B(z)$, which is a standard pedagogical issue. As for $\Psi_G(z)$, according to eq.ns~\eqref{E2} and \eqref{Sg2}, one gets, to first order in $\rho$:
\begin{equation}
\label{Psi}
\Psi_G(z,\:\rho)=-JN\left(z^2+2\rho\: z\right)-T\:\mathrm{ln}\left[2^{N+1}\int_{|z|}^{1}\mathrm{d}x\:e^{-\frac{N}{2}\phi(x)}\right]\:.
\end{equation}
First, let us study the perfectly symmetric system at $\rho=0$. The extremants of $\Psi_G(z,\:0)$ follow from the equation
\begin{equation*}
\label{eq.z2}
\frac{\partial\Psi_G(z,\:0)}{\partial z}=-2JN\:z+\mathrm{sign}(z)\:T\:e^{-\frac{N}{2}\phi\left(z\right)}\left[\int_{|z|}^{1}\mathrm{d}x\:e^{-\frac{N}{2}\phi(x)}\right]^{-1}=0\:,
\end{equation*}
which yields (see Appendix A):
\begin{subequations}
\label{zb2,zg2}
\begin{align}
|z|&=\frac{T}{2J}\mathrm{ln}\left(\frac{1+|z|}{1-|z|}\right)\quad&\text{for}\; |z|\gg\sqrt{\frac{1}{2N}}\label{zb2}\:,\\
\nonumber\\
&=\frac{T}{J\sqrt{2N\pi}}\equiv z_G\quad&\text{for}\; |z|\le\sqrt{\frac{1}{2N}}\:.\label{zg2}
\end{align}
\end{subequations}
In agreement with the rule that Gibbs and Boltzmann converge to the same results in large systems, Boltzmann's picture of Weiss ferromagnetism is recovered from eq.ns~\eqref{zb2,zg2}, in the limit $N\rightarrow\infty$. Indeed, one easily sees that in this limit $z=0$ is always a solution, while other two solutions $\pm z_{weiss}$ exist, from eq.n~\eqref{zb2}, provided  $T<T_c=2J$. The solutions $\pm z_{weiss}$ correspond to two minima of the Helmholtz free energy and yield the \emph{spontaneous magnetization}, below the Curie temperature $T_c$, where, according to Boltzmann, $z=0$ is a maximum (i.e., an unstable equilibrium state). Above $T_c$, the solution $z=0$ is the only minimum of Boltzmann's free energy, which marks the transition from the ferromagnetic to the paramagnetic phase. All this is displayed by the dashed line plots in Fig. 5.

\begin{figure}[htbp]
\begin{center}
\includegraphics[width=4in]{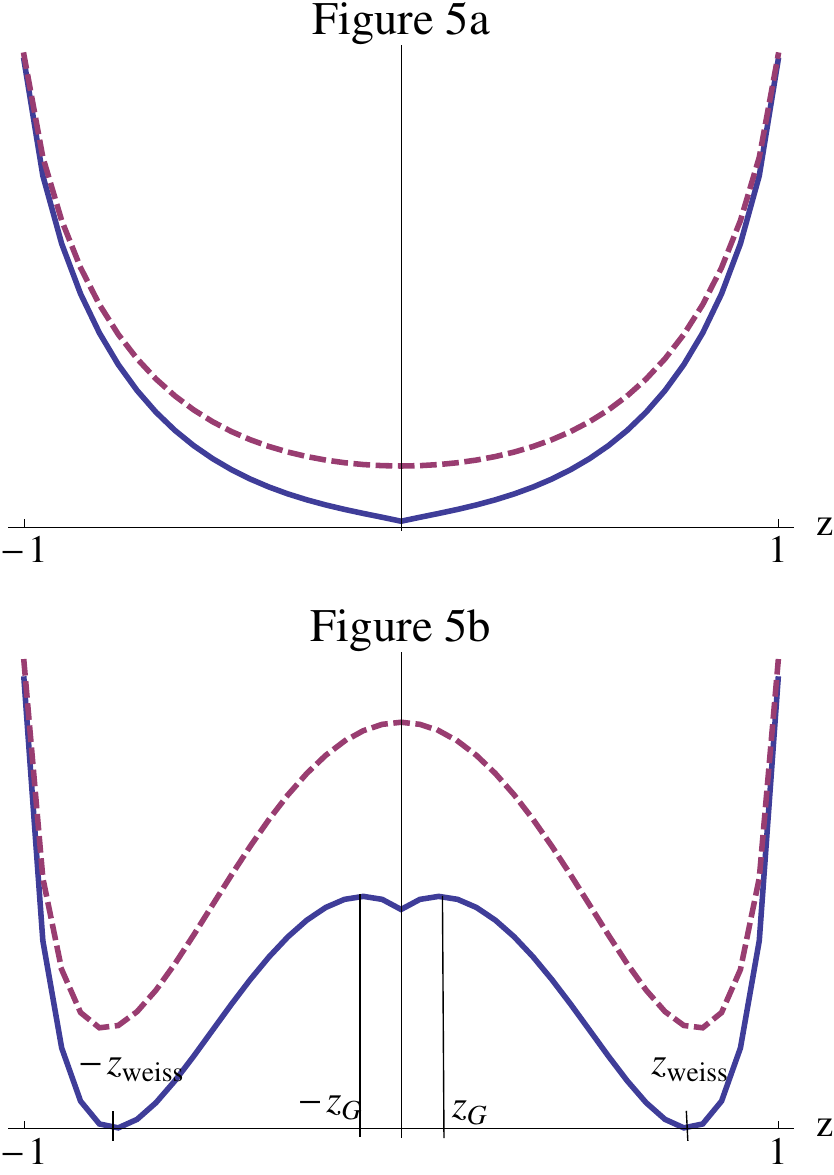}
\caption{\textbf{Gibbs (full lines) and Boltzmann (dashed lines) Helmholtz free energies as a function of $\mathbf{z}$ (arbitrary units, $N=200$).}\newline
(a) Paramagnetic regime ($T=1.25\times T_c$): $z=0$ is the only stable equilibrium state, both for Boltzmann and for Gibbs.\newline
(b) Ferromagnetic regime ($T=0.75\times T_c$): In addition to the two ferromagnetic states $\pm z_{weiss}$, Gibbs predicts the permanence of a stable paramagnetic state at $z=0$.\newline}
\label{default}
\end{center}
\end{figure}

In a finite system, Gibbs and Boltzmann pictures of Weiss ferromagnetism differ for small values of $|z|$. Indeed, for $T<T_M$ (recall Fig. 2), the solutions $\pm z_G=\pm T/\left(J\sqrt{2N\pi}\right)$ of eq.n~\eqref{zg2} correspond to two \emph{maxima} of $\Psi_G(z,\:0)$, while $z=0$ is always a \emph{minimum}, as can be seen by the full line plots in Fig. 5 and by noticing that (eq.n~\eqref{zg2}):

\begin{equation}
\label{cuspid}
\mathrm{lim}_{z\rightarrow0^{\pm}}\frac{\partial\Psi_G(z,\:0)}{\partial z}=\pm\:T\sqrt{\frac{2}{N\pi}}\:,  
\end{equation}

which shows that the minimum at $z=0$ corresponds to the terminal point of a downward cuspid, as sketched in Fig. 5.

In conclusion, the difference between Boltzmann and Gibbs in a finite-size ferromagnet, at zero external field, is that the former predicts an \emph{unstable} paramagnetic state $z=0$ in the ferromagnetic phase ($T<T_c$), while the latter predicts the persistence of a \emph{stable} paramagnetic phase, cohexisting with the two ferromagnetic states $\pm z_{weiss}$. In other words, Boltzmann predicts what is called a "spontaneous symmetry breaking" at $T=T_c$, in which an arbitrary small external field makes the system fall in one of the two ferromagnetic states $\pm z_{weiss}$, at $T<T_M$. Gibbs, in contrast, predicts a lower limit $B_m$ of the external field, below which the paramagnetic state persists, even in the ferromagnetic phase. The calculation of $B_m=2J\rho_m/\mu_0$ follows from eq.n~\eqref{Psi}, by looking for the small extremants in the presence of a small field:
\begin{equation*}
\frac{\partial\Psi_G(z,\:\rho)}{\partial z}=0\;\Rightarrow\;2(z+\rho)=\mathrm{sign}(z)\:z_G\quad\text{for } |z|\le\sqrt{\frac{1}{2N}}\:.
\end{equation*}
For this equation to have \emph{two} solutions, for both signs of $z$ (which preserves the minimum in between), it is necessary that
\begin{equation*}
|\rho|<\rho_m=\frac{z_G}{2}=\frac{T}{T_c\sqrt{2N\pi}}\:.
\end{equation*}
This shows that in a finite-size magnet, realized by elementary units of magnetic moment $\mu_0$, the symmetry breaking predicted by Gibbs is not "spontaneous", but involves a lower limiting value of the external magnetic field:
\begin{equation}
\label{Bmin}
|B|>B_m=\frac{T}{\mu_0\sqrt{2N\pi}}\quad (T<T_M)\:.
\end{equation}
vanishing with the inverse square root of $N$. The decrease of $B_m(N)$ with increasing $N$ corresponds to the "smoothing out" of the cuspid minimum. The $1/\sqrt{N}$-discrepancy between Boltzmann ($B_{min}=0$) and Gibbs ($B_{min}\propto1/\sqrt{N}$) is manifest from eq.n~\eqref{Bmin}.

The advantage of using a ferromagnetic material for testing the validity of Gibbs' entropy is that the recent developments of micro and nano-physics make it possible to prepare magnetic particles, containing controllable (and small) numbers of magnetic moments. Actually, size-effect measurements on such micro-magnets have become current in recent years \cite{Sal}. In the present case, one could measure $B_m$ with high sensistivity magnetometers, for different particles' size. If  SQUID magnetometers were to be used, at the largest operative temperature of about $10\:\mathrm{K}$, with $\mu_0\approx\mu_{Bohr}$, equation~\eqref{Bmin} yields $B_{min}\approx10^{-6}\mathrm{G}$ for a cube millimeter of magnetic sample, with a density $N/V\approx10^{23}\mathrm{cm}^{-3}$ typical of metals.

\section{Conclusions}

The contrast between Boltzmann's and Gibbs' approach to statistical thermodynamics is a long standing question, that comes to the light, time to time, since about one century, in different contexts, but with the same underlying \emph{leitmotiv}: which states are available to the system, and the way they must be counted. All this obviously reflects on the entropy. Dunkel and Hilbert, in ref.~\cite{DH} (see also ref.~\cite{HHD}), add a new element of discussion, by refreshing an almost forgotten definition of entropy, due to Gibbs (eq.n~\eqref{Sg}), that forbids negative absolute temperature (NAT), in contrast to the current expression~\eqref{Sb}, usually attributed to Boltzmann. The arguments in ref.~\cite{DH} have raised a discussion on the validity of eq.n~\eqref{Sg}, as an alternative to eq.n~\eqref{Sb} \cite{FW,Ual,DH',DH'',C,S}. The two expressions differ in the phase space regions uniformly occupied by the system: a surface of constant energy $E$, for Boltzmann; a volume containing \emph{all} states with energy less than $E$, for Gibbs. While Boltzmann's definition cannot apply but to an isolated, micro-canonical system, Gibbs entropy seems more appropriate for \emph{canonical} systems, whose energy can fluctuate, due to the heat exchanges with a thermal bath. This, however, contrasts with what is claimed in ref.s~\cite{DH, DH', C, HHD}. Hence, in addition to the question about which of the two expressions is correct, one should also explore the possibility that \emph{both} expressions are correct, but refer, respectively, to micro-canonical and canonical constraints (as stressed in ref.s~\cite{DH, DH', C, HHD}), or \emph{vice versa}.
  
Since Gibbs and Boltzmann entropies yield the same results in the thermodynamical limit $N\rightarrow\infty$, any possible \emph{measurable} consequence of what precedes results in a \emph{size effect} and is thereby far from easy to exploit. A superficial reading of eq.n (14) in ref.~\cite{DH} might lead to the conclusion that the smallness of the differences between eq.n~\eqref{Sb} and \eqref{Sg} vanish as $1/N$ (the inverse of an \emph{extensive} quantity like the heat capacity). If so, any experimental test, such as the "minimal quantum thermometer" suggested in ref.~\cite{DH}, would become extremely difficult, if not impossible.

The aim of the present work was to explore the possibility of concrete experimental procedures, deciding the winner (if any) of what we playfully called the Boltzmann \emph{vs} Gibbs match, or claiming that the two opponents actually play on different playgrounds, if Boltzmann and Gibbs entropies would result to refer to different constraints. 

The study of two-level systems, both in the form of gases (Section~\ref{2levgas}), and of interacting Ising spins (Sections~\ref{Weiss} and Appendix B), shows that the smallness factor to be accounted for is not $1/N$, but $1/\sqrt{N}$, which makes any experimental procedure much more feasible. In particular, it has been seen that some cube centimeters of $^3\mathrm{He}$, with density about $10^{23}\mathrm{cm}^{-3}$, at a temperature of tens Kelvins, under the action of a magnetic field of $10^{-7}\mathrm{G}$, could be an appropriate system for an experimental test, possibly revealing the most striking effect of Gibbs entropy, i.e. a \emph{thermally activated} overpopulation of the upper level. As a possible alternative, it is found that the effect supporting Gibbs' entropy in a Weiss ferromagnet (Ising model) should be a divergent heat capacity at a critical temperature $T_M$ (lower than the standard Curie temperature), in contrast to the finite discontinuity predicted by Boltzmann's picture at the Curie temperature $T_c$ (Figure 3). In a \emph{canonical} process (Section~\ref{Weiss2}), Gibbs' entropy would be also responsible for an anomalous persistence of the paramagnetic regime, below the critical temperature $T_M$, that should be destabilized by an external magnetic field proportional to $1/\sqrt{N}$ (eq.n~\eqref{Bmin}). This should be observed in canonical sub-millimetric metallic particles, with controllable number of interacting magnetic moments, under the action of weak external magnetic fields ($\approx10^{-6}\mathrm{G}$). 

The orders of magnitude involved in the measurements sketched above look fairly accessible to concrete experiments, which opens the possibility of an important advance in the thermodynamics of small systems, and in the genuine foundations of statistical thermodynamics.\newline
\\
\textbf{Post-amble}

As anticipated in the preamble, now I try to explain the reasons that lead one of the referees to reject the publication in \emph{PRE}. 

The referee denies that equation~\eqref{Tdef}, which is the cornerstone of all the question, is actually equivalent to the minimization of the Helmholtz free energy $\Psi$ with respect to, say, the magnetization (or any other state parameter), at fixed $V$ and $T$. The referee claims that, while this is true for Boltzmann's entropy, it might be false for Gibbs' entropy (giving no counter-example, however). In the lack of a first principle statistical derivation, says the referee, the minimization of $\Psi$ is an \emph{ad hoc} procedure, and any result supporting it might be a pure coincidence.  

As shown by what precedes, my point is that equation~\eqref{Tdef} is manifestly equivalent to find the \emph{extremants} of $\Psi$, which is nothing but a mathematical outcome. On using just equation~\eqref{Tdef} (and nothing else), I show that, in the case of Weiss ferromagnetism, some of those extremants behave as true thermal equilibrium values, while others behave like \emph{non equilibrium} values (see Fig.2), yielding, for example, a \emph{negative} heat capacity. Of course, the \textquoteleft good\textquoteright$\:$extremants and the \textquoteleft bad\textquoteright$\:$ones are the minima and the maxima of $\Psi$, respectively (see Fig.s 5). As far as I can see, the only problem with this picture is the \emph{opinion}, shared by many Gibbs' supporters, that Gibbs' entropy should apply to \emph{microcanonical} systems only. If so, the notion of Helmholtz free energy is questionable, since it is well known that minimizing $\Psi$ is a \emph{canonical} equilibrium procedure. This is aknowledged as a possible pending question to be explored, hopefully experimentally, which is right the aim of the present work: suggesting experimental tests to check the validity of two contrasting definitions of the entropy.

Paradoxically, the efforts made in the (vain) attempt to remove the referee's objections turned into some advantages. Thank to them, indeed, the work has been considerably improved. For example, the divergency of $C_G$ at $T_M$ (Fig. 3), that is an important physical aspect, had escaped to my attention in the first version.

\begin{appendices}
\numberwithin{equation}{section}
\section{Appendix}

For $N\gg1$, the contributes to the integral
\begin{equation*}
I\equiv\int_{-1}^{z}\:e^{-\frac{N}{2}\left[\phi(z)+a(x-z)+b(x-z)^2+\cdots\right]}\mathrm{d}x
\end{equation*}
in eq.n~\eqref{Tgas1} come from the minimum of $\phi(x)$ (positive and symmetric) in the integration interval $\left[-1,\;z(E)\right]$. So, $\phi(x)$ can be approximated by the first three terms in eq.n~\eqref{phi}, which yields:
\begin{subequations}
\begin{equation}
\label{AppI}
\:e^{\frac{N}{2}\phi(z)}I\approx \int_{-(1+z)}^{0}e^{-\frac{N}{2}\left(a\:y+b\:y^2\right)}\mathrm{d}y= \frac{1}{L_N}\int_{-(1+z)L_N}^{0}\:e^{-\left(Q^2+\Lambda_NQ\right)}\mathrm{d}Q\:,
\end{equation}
with
\begin{equation}
\label{L,Lambda}
 L_N\equiv\sqrt{\frac{Nb}{2}},\quad\Lambda_N\equiv a\sqrt{\frac{N}{2b}}\:.
\end{equation}
\end{subequations}
  
If $E<0$, one has $z(E)<0$ in turn (eq.n~\eqref{z}) and $\Lambda_N<0$. If $\left|\Lambda_N\right|$ is large, $Q^2$ can be neglected in the exponent of the integrand in eq.n~\eqref{AppI}, which yields:
\begin{equation}
\label{AppE<0}
e^{\frac{N}{2}\phi(z)}I\approx\frac{1}{L_N}\int_{-\infty}^{0}\:e^{-\Lambda_NQ}\mathrm{d}Q=\frac{2}{|a(z)|N}\quad\text{for }\left|\Lambda_N\right|\gg1\:.
\end{equation}
On replacing expression \eqref{AppE<0} in eq.n~\eqref{Tgas1}, one gets eq.n~\eqref{E<0}, with the same validity condition (recall eq.ns~\eqref{phi} and \eqref{L,Lambda}).

If $E>0$ ($z,\:\Lambda_N>0$), the integral in eq.n~\eqref{AppI} can be calculated by completing the square in the exponent:
\begin{equation}
\label{AppE>0}
e^{\frac{N}{2}\phi(z)}I\approx\frac{e^{\Lambda_N^2/4}}{L_N}\int_{-\infty}^{\Lambda_N/2}\:e^{-R^2}\mathrm{d}R=\frac{e^{\Lambda_N^2/4}\sqrt{\pi}}{2L_N}\left[1+\mathrm{erf}\left(\frac{\Lambda_N}{2}\right)\right]\:. 
\end{equation}
If, in particular, one takes $z\ll1$, equations~\eqref{AppE>0}, \eqref{phi} and \eqref{L,Lambda} lead to the expression in eq.n~\eqref{E>0}.

The integral
\begin{equation}
\int_{|z|}^{1}\:e^{-\frac{N}{2}\left[\phi(z)+a(x-z)+b(x-z)^2+\cdots\right]}\mathrm{d}x\:,
\end{equation}
in eq.n~\eqref{eq.z2} can be calculated with the same method, for $\left|\Lambda_N\right|\gg1$ (eq.n~\eqref{zb2}) and $\left|\Lambda_N\right|\le1$ (eq.n~\eqref{zg2}).

\end{appendices}

\textbf{Aknowledgments}\newline
The author id grateful to dr R. Costa for his bibliographical contribution and for useful discussions, and to prof L. Del Bianco for her explanations about SQUIDs' and magnetic moment measurements in general.
\end{document}